\documentclass[12pt]{article}
\usepackage{amssymb}
\usepackage{amsmath}
\usepackage{graphicx}

\renewcommand{\vec}[1]{{\bf #1}}

\def\beq{\begin{eqnarray}}
\def\eeq{\end{eqnarray}}
\def\ln{\,\mbox{ln}\,}

\def\ga{\gamma}\def\de{\delta}

\def\ep{\epsilon}

\def\la{\lambda}

\def\pa{\partial}

\def\La{\Lambda}

\def\Om{\Omega}

\input{psfig}

\begin{document}

%
\begin{center}
{\Large\sc Simple Cosmological Model with Relativistic Gas}
\vskip 5mm
{\small \bf Guilherme de Berredo-Peixoto}
\footnote{E-mail address: guilherme@fisica.ufjf.br},
\qquad
{\small \bf Ilya L. Shapiro} \footnote{On leave
from Tomsk State Pedagogical University, Russia. E-mail address:
shapiro@fisica.ufjf.br},
\qquad
{\small \bf Fl\'avia Sobreira}
 \footnote{E-mail address: flavia$\_$sobreira@yahoo.com.br}
\vskip 3mm
{\small\sl Departamento de F\'{\i}sica -- ICE,
Universidade Federal de Juiz de Fora,} \\
{\small\sl Juiz de Fora, CEP: 36036-330, MG,  Brazil}
\end{center}
\vskip 6mm
\centerline{\uppercase{abstract}}
\vskip 2mm

\begin{quotation}
\noindent
We construct simple and useful approximation for the
relativistic gas of massive particles. The equation of
state is given by an elementary function and admits
analytic solution of the Friedmann equation, including
more complex cases when the relativistic gas of massive
particles is considered together with radiation or with
dominating cosmological constant. The model of relativistic
gas may be interesting for the description of primordial
Universe, especially as a candidate for the role of a
Dark Matter.
\vskip 2mm

\end{quotation}

\vskip 4mm                                                  %
\section{\large\bf  Introduction}                           %

$\,$ \quad
During many years most of the works about cosmological
solutions assumed stationary equation of state for the matter
or vacuum sources of the Friedmann equations. The conventional
equation of state is a linear relation between pressure
and energy density $\,P=w\rho$.
Different values of the parameter $\,w\,$ correspond to
different kinds of sources. For example, $\,w=-1\,$
holds for the cosmological constant (vacuum energy),
$\,w=0\,$ for the pressureless (dust-like) matter,
$\,w=1/3\,$ for the radiation. Recently, due to the growing
amount and quality of the observational/experimental
cosmological data, theoretical considerations involved
more complicated equations of state, with $\,w\,$
depending on time or/and energy density. Perhaps the
first examples of this kind were related to inflation,
where the variable vacuum energy density was introduced due
to the electroweak phase transition \cite{Guth} and
the non-trivial inflaton potential \cite{Linde}.

Theoretical investigations of the models with variable
vacuum energy were fueled recently by the new precise
measurements of the expansion rate of the Universe from
the type Ia supernovae experiments \cite{SN} and also from
the cosmic microwave background radiation \cite{CMB}.
The existing data indicate
that the Universe is mainly composed by the non-luminous
sources, such as Dark Matter, responsible for 20-30\%
of the overall energy density balance, and the Dark Energy,
responsible for 65-75\%. One of the main candidates to the
role of the most of the Dark Matter is a gas of weakly
interacting massive particles, e.g., the ones corresponding
to the broken supersymmetry. Let us notice that recent 
theoretical and phenomenological considerations of the 
supersymmetric neutralino do not rule out the
light DM options \cite{boehm,baier} (see \cite{fayet,bel} for
the review). The main candidate to be Dark
Energy is the cosmological constant with equation of state
$\,w=-1$. The anthropic considerations show that the total
value of the cosmological constant {\it should be} positive
and in fact close to the observed one \cite{weinberg,vilenkin}.

However, since nobody can guarantee that the vacuum energy
is red-shift independent, it is quite
natural to meet a variety of alternative models for the
Dark Energy, such as quintessence \cite{quintess},
Chaplygin gas \cite{Chapl} and the low-energy
renormalization group running of the proper cosmological
constant \cite{CC-fit}. Most of these models lead to a
variable $\,w\,$ in the equation of state for the vacuum
energy. This effect is achieved either by postulating this
equation of state in the case of Chaplygin gas or by
postulating a properly chosen quintessence potential.

There is another possibility to meet the equation of
state with a variable $\,w$, depending on the energy density
and therefore on the red-shift. In the present article
we shall consider the Universe filled by the ideal gas
of relativistic massive particles\footnote{The cosmological
model taking into account the relativistic effects related
to the peculiar velocities of galaxies has
been developed in \cite{soleng}.}. This model may have
interesting applications, e.g., in the
early radiation-dominated Universe we can consider
relativistic gas of massive particles as a model for
the hot matter content. Furthermore, relativistic effects
may be, in principle, relevant for the Dark Matter problem.
Since we do not know exactly from what the Dark Matter
is done, any possibility here deserves
careful exploration. Indeed, when assuming that the
Dark Matter is a relativistic gas, we suppose that
it is composed (or has been composed in the earlier epochs)
from the relatively light, massive particles weakly
interacting with the baryonic matter and radiation.

The equation of state for the relativistic ideal gas is
known for a long time \cite{Juttner} (see also, e.g.,
\cite{Pauli})\footnote{
The relativistic version of Maxwell distribution follows
from the corresponding generalization of the Boltzmann
$H$-theorem (see, e.g. \cite{Kremer}).}. The relation
between pressure and energy density involves modified
Bessel functions. Obviously, this form of
equation of state in not very useful for cosmological
applications. At the same time one can considerably
simplify the cosmological model with relativistic gas
without losing much of the physical sense. In order to
do so we shall assume that, instead of following Maxwell
distribution, all particles have equal kinetic energy.
Below we call the model of relativistic gas of massive
particles
with equal energies the {\it reduced relativistic gas}.

The ``defect'' in the equation of state which follows
from the assumption of equal energies is not very significant.
The numerical comparison with the Maxwell distribution is
presented in the Appendix. It turns out that the difference
between the two distributions does not exceed 2.5\% even
in the low-energy
region, being negligible for the ultrarelativistic gas.
Let us remember that the Maxwell distribution is also just
an approximation to the real situation. For example, when
considering the Maxwell distribution for the identical
massive particles in the Early Universe we are
disregarding interactions between these particles and
radiation, and also differences between the masses
of different kinds of particles.

At the same time, the model of reduced relativistic gas
provides a great
advantage for cosmological applications. Starting from
the reduced equation of state one can integrate the Friedmann
equation analytically, leading to a nice and simple
cosmological model interpolating between radiation-dominated
and  matter-dominated epochs of the Universe. In the
present paper we shall develop this model and also consider
more complicated cases of the Universe filled by reduced
relativistic gas plus radiation and of the Universe dominated
by the cosmological constant where matter content is the
reduced relativistic gas.

The paper is organized as follows. In the next Section
we shall briefly describe the conventional model of
relativistic gas and our model of the reduced relativistic
gas. In Section 3 we use the equation of state for the
reduced model and obtain the scale dependence for the
energy density and pressure. Section 4 is devoted to the
solution of the cosmological model with  the reduced
relativistic gas for
several interesting particular cases. In section 4 we
draw our conclusions.

\section{\large\bf Reduced model for relativistic gas}  %

$\,$ \quad
Consider a single relativistic particle with the rest
mass $\,m\,$ in a volume $\,V$. The dispersion
relation for this particle has standard form
\beq
\ep^2 - c^2 {\vec p}^2 = m^2 c^4
\,,\quad \mbox{where} \quad
{\vec p}=\frac{m{\vec v}}{\sqrt{1-v^2/c^2}}\,.
\label{dispersion}
\eeq
An elementary
consideration shows that the time average of the pressure
produced by the particle on the walls of the vessel is
\beq
P = \frac{1}{3\,V}\,\cdot\,\frac{m v^2}{\sqrt{1-v^2/c^2}}\,.
\label{pressure}
\eeq
For the gas of $\,N\,$ such particles with equal kinetic
energies $\,\ep$, we arrive at the following equation of
state
\beq
P = \frac{\rho}{3}\,\cdot\,
\left[\,1-\Big(\frac{m c^2}{\ep}\Big)^2\,\right]
\,,\quad \mbox{where} \quad
\rho=\frac{N\,\ep}{V}
\label{state 1}
\eeq
is the energy density. Let us notice that $\,w=P/\rho$
tends to $1/3$ in the ultra-relativistic limit
$\ep\to \infty$ and to zero in the non-relativistic
limit $\ep\to mc^2$. It proves useful to introduce the
following new notations: the density of the rest energy
of the particles at the initial moment
$\,\rho_1 = Nmc^2/V_0$, where $\,V_0\,$ is
some fixed initial value of the volume, and also
$\,\rho_d = \rho_d(V) = \rho_1 V_0/V = Nmc^2/V$, which
shows how
the same density evolves with the change of the volume.
Using these notations we can cast the equation
(\ref{state 1}) into the form
\beq
P = \frac{\rho}{3}\,\cdot\,
\left[\,1\,-\,\frac{\rho_d^2}{\rho^2}
\,\right] \,.
\label{state 2}
\eeq

In order to understand better the difference between
these formulas and the "correct" ones, let us consider
the Maxwell distribution for the ideal gas of massive
particles. The statistical integral for a single particle
is given by the expression
\beq
Z = \int e^{-\ep/kT}\,d^3p\,d^3q
\,=\,4\pi\,m^2c\,V\,\cdot\,K_2\Big(\frac{mc^2}{kT}\Big)\,,
\label{M state 1}
\eeq
where $\,K_\nu(x)\,$ is a modified Bessel function of
index $\,\nu$.
The equation of state for the
gas of $N$ particles can be derived in a standard way
\beq
PV\,=\,kT\,N\,\,\Big(\frac{\pa \ln Z}{\pa \ln V}\Big)_T
= N\,kT\,,
\label{M state 2}
\eeq
while the average energy of the particle is
\beq
{\bar \ep}
\,=\, \frac{1}{Z}\,\int \,e^{-\ep/kT}\,\ep \,d^3p\,d^3q
\,=\, mc^2\,\frac{K_3(mc^2/kT)}{K_2(mc^2/kT)} \,-\, kT\,.
\label{M state 3}
\eeq
The energy density $\,\rho=N {\bar \ep}/V\,$
and pressure $\,P\,$ are related by an implicit functional
dependence (\ref{M state 2}) and
(\ref{M state 3}), which has to be compared with the
formula (\ref{state 2}) for the reduced gas case. This
comparison will be performed numerically in the Appendix,
where we show that (\ref{state 2}) is an excellent
approximation to the relations (\ref{M state 2}),
(\ref{M state 3}). For a while we conclude this section
by an obvious observation that
(\ref{state 2}) is much simpler than (\ref{M state 2})
plus (\ref{M state 3}).

\section{\large\bf Scale dependence in the reduced model}  %

$\,$ \quad
Let us use the equation of state (\ref{state 2}) for the
reduced relativistic gas and find how the energy density
depends on the volume under adiabatic expansion.
For this end we replace the equation (\ref{state 2})
into the conservation law
\beq
- \,\frac{dV}{V}\,=\,\frac{d\rho}{\rho + P}\,.
\label{scale 1}
\eeq
In the cosmological setting the last equation implies
that the reduced relativistic gas does not exchange
energy with other entities like with radiation in the
Early Universe or with baryonic matter and
vacuum energy in the case when the reduced relativistic
gas is considered as a model for the Dark Matter.

The differential equation (\ref{scale 1}) can be
easily solved. The solution has the form
\beq
\rho(V)\,=\,\left[\,\rho_1^2\,\Big(\frac{V_0}{V}\Big)^2
\, + \, \rho_2^2\,\Big(\frac{V_0}{V}\Big)^{8/3}
\,\right]^{1/2}\,,
\label{scale 2}
\eeq
where the initial condition has been defined as
\beq
\rho(V_0)\,=\,\left[\,\rho_1^2 \,+\, \rho_2^2
\,\right]^{1/2}\,.
\nonumber
\eeq
Remember that $\,\rho_1=Nmc^2/V_0\,$ is the rest energy
density at the initial point $\,V=V_0$. Hence the second
component $\,\rho_2\,$ can be interpreted as the energy
density of the radiation component of the reduced
relativistic gas. However the total energy density
is not a simple sum of the two components but the square
root of the sum of their squares. The non-relativistic or
ultra-relativistic limits are achieved when one takes,
correspondingly, $\,\rho_2=0\,$ or $\,\rho_1=0$. It is
easy to see that the expression (\ref{scale 2}) provides
correct scaling laws in these two cases.

For the sake of cosmological applications it is better
to express the energy density as a function of the conformal
factor $\,a$, where $\,(a/a_0)^3=V/V_0$. Then we
arrive at the formula
\beq
\rho(a)\,=\,\left[\,\rho_1^2\,\Big(\frac{a_0}{a}\Big)^6
\, + \, \rho_2^2\,\Big(\frac{a_0}{a}\Big)^8
\,\right]^{1/2}\,,
\label{scale 3}
\eeq
with the initial condition
$\,\rho(a_0)=\left[\rho_1^2+\rho_2^2\right]^{1/2}$. The
density of the rest mass behaves like
$\,\rho_d(a)=\rho_1 (a_0/a)^3$. The
scale dependence of the pressure and parameter $\,w\,$
are given by the eq. (\ref{state 2}) and the equation
\beq
w = \frac{1}{3}\,\cdot\,
\left[\,1\,-\,\frac{\rho^2_d(a)}{\rho^2(a)}\,\right]\,.
\label{w}
\eeq
It is easy to see that the relations (\ref{state 2}) and
(\ref{w}) predict dust-like (pressureless) scaling in the
limit $\,a\to \infty\,$ and the radiation-like scaling
$\,w \approx 1/3\,$ in the limit $\,a\to 0$. Hence
our model of reduced relativistic gas can be regarded as
an interpolation model between the radiation-dominated
and matter-dominated evolutions. If the Universe, from
the very beginning, were filled by \ {\it ideal} \ hot
gases of massive particles, there would not be nuclear
reactions and the scale behaviour of the energy density
would be close to (\ref{scale 3}).

\section{\large\bf Solving the Friedmann equation}      %

$\,$ \quad
Consider the cosmological model of the Universe
filled by the reduced relativistic gas (\ref{scale 3}).
For the sake of generality, let us start by formulating
the Friedmann equation for an arbitrary $\,k\,$
and also include vacuum energy density,
$\,\rho_\La=\La/8\pi G\,$, and the radiation energy
density $\,\rho_{r}(a)=\rho_{r0}/a^4$. In what follows
we set $\,a_0=1$. The equation of interest has the form
\beq
\Big(\frac{{\dot a}}{a}\Big)^2 + \frac{k}{a^2}
= \frac{8\pi G}{3}
\,\big[\,\rho(a)+\rho_\La+\rho_r(a)\,\big]\,,
\label{F}
\eeq
where $\,\rho(a)\,$ is given by (\ref{scale 3}). One
can easily add other matter sources and solve
the resulting equation following the examples which
will be elaborated below.

\subsection{\bf Pure reduced gas model}                 %

As a first particular case we shall integrate eq. (\ref{F}) for
the pure reduced relativistic gas model, $\,k =\rho_\La=\rho_r=0$.
After introducing a new variable $\,x=a^2\,$ the equation becomes
\beq
{\dot x}^2 = \frac{32\pi G\,\rho_1}{3}\,\sqrt{x+b}\,, \qquad
\mbox{where}\qquad b=\frac{\rho_2^2}{\rho_1^2}\,. \label{F 1} \eeq
The solution of this equation has the form \beq \left(\,a^2 +
\frac{\rho_2^2}{\rho_1^2}\,\right)^{3/4} =\sqrt{6\pi
G\rho_1}\,\cdot\,t\,.
\label{F 2}
\eeq
 The last expression shows,
again, that our model interpolates between the usual
matter-dominated and radiation-dominated FRW solutions. In the
non-relativistic case, $\,\rho_2=0$, we obtain directly the
standard behaviour $\,a(t) \sim t^{2/3}$. The ultrarelativistic
regime can not be considered in a direct way, because the limit
$\,\rho_1=0\,$ is singular. Let us assume $\,\rho_1 \ll \rho_2\,$
and expand until the lowest nontrivial order in the ratio
$\,(\rho_1/\rho_2)$
\beq
\Big(\frac{\rho_2}{\rho_1}\Big)^{3/2}\,\,
\left[1\,+\,\frac34\,\Big(\frac{\rho_1}{\rho_2}\Big)^2
\,a^2\right] \,=\,\sqrt{6\pi G\rho_1}\,\,t\,.
\label{F 3}
\eeq
The last expression is nothing else but the usual radiation-dominated
solution with the shifted time variable
\beq
a(t)\,=\,
\left(\frac{32\pi G\rho_2}{3}\right)^{1/4}\,\sqrt{t-t_0} \,,\qquad
\mbox{where} \qquad t_0 = \frac{\rho_2^{3/2}}{\sqrt{6\pi
G}\cdot\rho_1^2}\,. \label{F 4}
\eeq
Indeed, such time shifts are
irrelevant and we will not pay attention to them in what follows.

\begin{figure}[tb]
\vskip -0.5cm
\centerline{\resizebox*{8.0 cm}{!}{\includegraphics{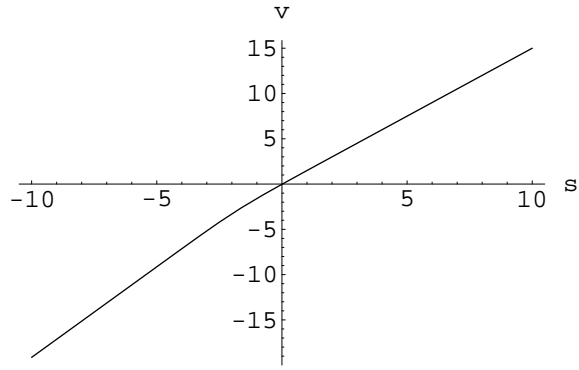}}}
\begin{quotation}
\caption{\sl The log-log plot for the transition between the
radiation-dominated and matter-dominated epochs in the pure
reduced relativistic gas model.}
\end{quotation}
\end{figure}

In order to illustrate the behaviour of the conformal
factor (\ref{F 2}) in the transition period, let us choose
the logarithmic parametrization for both time and
conformal factor. For the sake of simplicity we fix $t_0$
according to (\ref{F 4}) and rewrite (\ref{F 2}) in the
form
\beq
e^v\,=\,\ln \left[
\Big(e^{2s} + \frac{\rho_2^2}{\rho_1^2}\Big)^{3/4}
- \Big(\frac{\rho_2}{\rho_1}\Big)^{3/2} \right]\,,
\label{dubina}
\eeq
where $\,e^v = \sqrt{6\pi G\rho_1}\,t\,$ and
$\,a=e^{s}$.

The plot of $\,v\,$ versus $\,s\,$ depends on the
value of a single parameter $\,\rho_2/\rho_1\,$ and
in all cases it clearly demonstrates the transition between
the two linear asymptotic regimes. We present an example
for $\,\rho_2/\rho_1=1\,$ at the Figure 1. The reader
can easily achieve similar plots for other values of this 
parameter, also in the presence of a radiation
content. The solution for the last case is presented in the 
next subsection.

\vskip 4mm
\subsection{\bf Radiation dominated epoch}              %

Consider the reduced relativistic massive gas in the
radiation-dominated epoch. According to the known
estimates, cosmological constant and space curvature
are not very relevant in this case \cite{Bludman,Weinberg}
and we can safely set $\,k =\rho_\La=0$.

The solution of the Friedmann equation (\ref{F}) can be
written as (parameter $\,b\,$ is defined in (\ref{F 1}))
\beq
\frac43\,
\left[\sqrt{a^2+b}+\frac{\rho_{r0}}{\rho_1}\right]^{3/2}
\,-\,\frac{4\rho_{r0}}{\rho_1}\,
\left[\sqrt{a^2+b}+\frac{\rho_{r0}}{\rho_1}\right]^{1/2}
\,=\,
\left(\frac{32\pi G\rho_1}{3}\right)^{1/2}\,t\,.
\label{r 1}
\eeq
The last relation represents exact solution, but it is
too complicated for the qualitative analysis. Let us
consider the special situation when the radiation energy
density is strongly dominated and the effect of the
reduced relativistic massive gas is a small correction
to the $\,a\sim t^{1/2}\,$ law. Our purpose is to
evaluate this correction. For this end one has to expand
the expression (here $\,\ga=1/2\,$ or $\,3/2$)
\beq
\left[\sqrt{a^2+b}+\frac{\rho_{r0}}{\rho_1}\right]^{\ga}
\,=\,
\left(\frac{\rho_{r0}}{\rho_1}\right)^{\ga}\,
\left[\,1\,+\,\frac{\rho_1}{\rho_{r0}}\,
\sqrt{a^2+b}\,\right]^\ga
\label{r 2}
\eeq
until the third order in the small parameter
$\,\frac{\rho_1}{\rho_{r0}}\,\sqrt{a^2+b}\,$.

\begin{figure}[tb]
\vskip -0.5cm
\centerline{\resizebox*{8.0 cm}{!}{\includegraphics{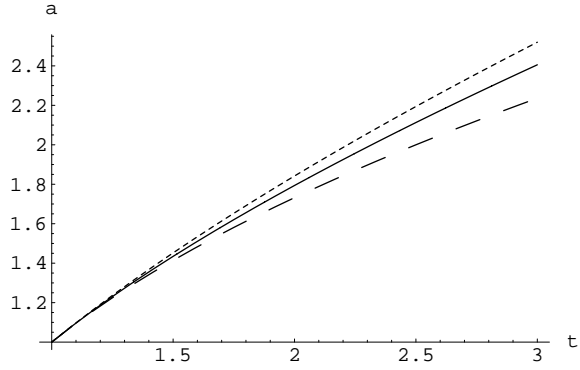}}}
\begin{quotation}
\caption{\sl The plot for the conformal
factor $a(t)$ in the following three cases: pure radiation
$a(t)\sim t^{1/2}$ is represented by the dashed line, pure
dust-like case by the dots line,
the case of pure reduced relativistic
gas is represented by the continuous line.}
\end{quotation}
\end{figure}

The result has the form
\beq
a^2 \,-\,\frac13\,\left(\frac{\rho_1}{\rho_{r0}}\right)\,
\left[\,a^2\,+\,b\,\right]^{3/2}
\,=\,
\left(\frac{32\pi G\rho_{r0}}{3}\right)^{1/2}\,t\,,
\label{r 3}
\eeq
where we disregard the initial value $t_0$. It is easy to
see that the effect of the reduced relativistic gas is to
accelerate the expansion of the Universe compared to the
pure radiation content. The effect of relativistic gas of
massive particles is weaker that the one coming from the
dust-like matter with the same energy density. The
illustrative plots of the reduced relativistic gas compared
to radiation and dust cases are presented at the Figure 2.

%
\subsection{\bf Cosmological constant dominated epoch}  %

The next relevant particular case is the
reduced relativistic gas as a model for the Dark Matter,
in a Universe dominated by the vacuum energy.
After the usual change of variable, $\,x=a^2$,
we arrive at the solution in the form of an integral
\beq
\int \frac{dx}{\left[\rho_\La x^2
+ \rho_1 \sqrt{x+b}\right]^{1/2}}
\,=\, \sqrt{\frac{32\pi G}{3}}\,t\,,
\label{c 1}
\eeq
with $\,b\,$ defined in (\ref{F 1}).
This integral is rather complicated and difficult to
evaluate analytically. Hence we can either apply the
numerical method or use the dominant role of the
vacuum energy density. The results of numerical
analysis for $\,\Om_\La=0,7\,$ and $\,\Om_M=0,3\,$
are shown at the Figure 3.
The presence of the reduced relativistic massive gas
results in the slower expansion of the Universe compared
to the pure cosmological constant case.

%
\begin{figure}[tb]
\vskip -0.5cm
\centerline{\resizebox*{8.0 cm}{!}{\includegraphics{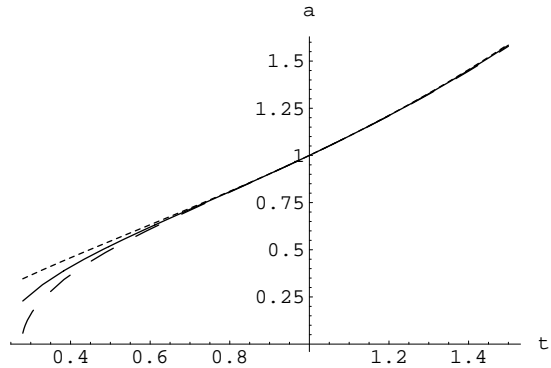}}}
\begin{quotation}
\caption{\sl The plot of conformal factor $\,\,a(t)\,\,$ for the
following three cases: cosmological constant plus ordinary
pressureless matter (points line), cosmological constant
plus pure radiation (dashed line) and cosmological constant plus
reduced relativistic gas (continuous line). One can see that the
effect of the reduced relativistic gas is intermediate between the
ones of the dust-like and radiation-like matter contents. At later
times all three curves converge and only the cosmological constant
remains relevant.}
\end{quotation}
\end{figure}
%


Consider an approximate analytical solution based on
the assumption of the dominant contribution of the
vacuum energy density. We rewrite
the integral in the {\it l.h.s.} of eq. (\ref{c 1})
in the form
\beq
\frac{1}{\sqrt{\rho_\La}}\,
\int \frac{dx}{x\,\sqrt{1+\ga}}
\,,\qquad\mbox{where}\qquad
\ga = \frac{\rho_1}{\rho_\La}\,\frac{\sqrt{x+b}}{x^2}
\nonumber
\eeq
and expand into power series in the small parameter
$\ga$. Then the integration becomes trivial and taking
the lowest nontrivial order into account we arrive at
the solution
\beq
\ln a \,+\,X[a]\,=\,\la\,t\,,
\label{c 2}
\eeq
where
\beq
X[a]\,=\,\frac{\rho_1}{8\,\rho_\La}\,
\left\{\,\frac{\sqrt{a^2+b}}{a^4}
+ \frac{\sqrt{a^2+b}}{2b\,a^2}
+ \frac{1}{4\,b^{3/2}}\,
\ln \Big(\,\frac{\sqrt{a^2+b}
-\sqrt{b}}{\sqrt{a^2+b}+\sqrt{b}}\,\Big)
\right\}
\label{c 22}
\eeq
is a small term and $\,\la = \sqrt{{8\pi G\rho_\La}/{3}}$.
One can easily find an approximate explicit formula for
$a(t)$ starting from the expression
\beq
a(t)\,=\,\big[1+f(t)\big]\cdot \,e^{\la t}
\,,\qquad \left|f(t)\right| \ll 1\,,
\label{c 23}
\eeq
and using the smallness of \ $X[a]$.
\ Replacing \ (\ref{c 23}) \
into \ (\ref{c 2}) \ we arrive at the solution
$\,\,f(t)=-X[e^{\la t}]$. This expression can be replaced
into (\ref{c 23}) to give
\beq
a(t)\,=\,e^{\la t}\,\left(\,1-X[e^{\la t}]\,\right)\,.
\label{c 24}
\eeq
Qualitatively the behaviour of $a(t)$ fits with the numerical
analysis. The presence of reduced relativistic gas slows
the acceleration of the Universe caused by $\,\rho_\La$.

\section{\large\bf Conclusions}                         %

$\,$ \quad
We constructed a simple and useful model of reduced
relativistic gas of massive particles, starting from the
``primitive distribution'' for kinetic energies.
Our model is a very good approximation to the
much more complicated Maxwell distribution (see Appendix).
The main advantage of the reduced model is that it admits
analytical derivation of the dependence $\,\rho(a)\,$ in
terms of elementary (and very simple) functions. Moreover,
one can easily integrate the Friedmann equation for the
Universe filled by the reduced relativistic gas, again
the solution is given by elementary function. In this
way we arrive at the interpolation between the cosmological
solutions for the matter-dominated and radiation-dominated
cases.

In the more complex situation, when the sources of the
Friedmann equation include also radiation or the cosmological
constant term, one can obtain the solution in the form
of an integral. The integration can be performed
numerically or analytically using the assumption of
radiation dominance or cosmological constant dominance.
In the last case the reduced relativistic gas
can be viewed as a model for the Dark Matter.

The model of reduced relativistic gas may be, in principle,
testable. Imagine the Dark Matter is composed by the
relatively light, weakly interacted massive particles
which have, at present, $\,w\approx 0\,$ equation of
state. One can not rule out the possibility that the
relativistic effects of these particles were relevant
in the earlier epochs of the Universe, e.g. in the
structure formation period. Then, some the traces of
these relativistic effects may be eventually found in
the precise CMB measurements. Therefore it would be
interesting to investigate density
and metric perturbations in this model. We postpone this
problem for the future work.

\section*{\large\bf Appendix}                            %

The purpose of this Appendix is to compare the results
(\ref{M state 2}) and (\ref{M state 3}) of the Maxwell
distribution for the relativistic gas of massive particles
and the corresponding relation (\ref{state 2}) for
the reduced relativistic gas model.
In the formulas (\ref{M state 2}) and (\ref{M state 3})
the temperature $\,kT\,$ plays the role of parameter
and we are in fact interested only in the dependence
between $\,P\,$ and $\,\rho$. Hence we solve
(\ref{M state 2}) with respect to pressure and replace
it into (\ref{M state 3})
\beq
\rho_M(P) = \frac{K_3(\rho_d /P)}{K_2(\rho_d/P)}
\,\,\rho_d \,-\,P\,,
\label{Maxwell}
\eeq
where the subscript $M$ indicated Maxwell distribution.

It is easy to see that the non-relativistic limit $P\to 0$
and the ultrarelativistic limit $P\to\rho/3$ of the last
relation coincide with the one for the reduced model
(\ref{state 2}). In order to compare the two expressions
numerically at the intermediate scales, let us rewrite
(\ref{state 2}) in the form similar to (\ref{Maxwell})
\beq
\rho(P) \,=\, \frac{3}{2}\,P\,+\,
\sqrt{\frac{9P^2}{4} + \rho_d^2}\,.
\label{similar}
\eeq
After assigning numerical value to $\rho_1$
(we set $\,\rho_1=1$, but the result does not depend
on this choice.), one can plot the relative
deviation
\beq
\de_\rho\,=\,\frac{| \rho - \rho_{M}|}{\rho_{M}}
\nonumber
\eeq
versus $\,P$. The plot for this dependence in presented
at the Figure 4. It is easy to see that the relative
deviation $\,\de_\rho\,$ achieves its maximum of about 2.5\%
in the non-relativistic region and becomes completely
negligible in the higher energy region.

%
\begin{figure}[tb]
\vskip -0.5cm
\centerline{\resizebox*{8.0 cm}{!}{\includegraphics{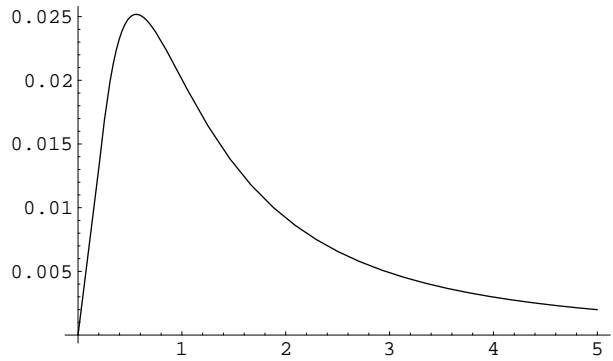}}}
\begin{quotation}
\caption{\sl Plot of \ $\de_\rho$ \ for the \ $\rho_0 = 1$ \ case.
The maximum discrepancy is about $\,\,2.5\%$ .}
\end{quotation}
\end{figure}

\vskip 4mm

\noindent
{\large\bf Acknowledgments.}
I.Sh. is indebted to A. Belyaev and J. Fabris for 
useful discussions.
The work
of the authors has been supported by the scholarship from
CNPq (F.S.), the fellowship from FAPEMIG (G.B.P.) and the
grants from CNPq and FAPEMIG (I.Sh.).

\vskip 4mm

\end{document}